\documentclass[aps,prl,reprint,noeprint,showpacs,showkeys,superscriptaddress,longbibliography,floatfix]{revtex4-1} 
\usepackage[utf8]{inputenc}
\usepackage{cmap} 
\usepackage[T1]{fontenc}
\usepackage{amsmath,amssymb,mathtools}

\usepackage{graphicx}
\graphicspath{.}

\begin{document}
 
\title{Aging in the long-range Ising model}
\author{Henrik Christiansen}
\email{henrik.christiansen@itp.uni-leipzig.de}
\affiliation{Institut f\"ur Theoretische Physik, Universit\"at Leipzig, IPF 231101, 04081 Leipzig, Germany}
\author{Suman Majumder}
\email{suman.majumder@itp.uni-leipzig.de}
\affiliation{Institut f\"ur Theoretische Physik, Universit\"at Leipzig, IPF 231101, 04081 Leipzig, Germany}
\author{Malte Henkel}
\email{malte.henkel@univ-lorraine.fr}
\affiliation{Laboratoire de Physique et Chimie Th\'eoriques (CNRS UMR 7019), Universit\'e de Lorraine Nancy, 54506 Vand{\oe}uvre-l\`es-Nancy Cedex, France}
\affiliation{Centro de F\'isica Te\'orica e Computacional, Universidade de Lisboa, 1749-016 Lisboa, Portugal}
\author{Wolfhard Janke}
\email{wolfhard.janke@itp.uni-leipzig.de}
\affiliation{Institut f\"ur Theoretische Physik, Universit\"at Leipzig, IPF 231101, 04081 Leipzig, Germany}
\date{\today}
 
\begin{abstract}
  The current understanding of aging phenomena is mainly confined to the study of systems with short-ranged interactions.
  Little is known about the aging of long-ranged systems.
  Here, the aging in the phase-ordering kinetics of the two-dimensional Ising model with power-law long-range interactions is studied via Monte Carlo simulations.
  The dynamical scaling of the two-time spin-spin autocorrelator is well described by simple aging for all interaction ranges studied. 
  The autocorrelation exponents are consistent with $\lambda=1.25$ in the effectively short-range regime, while for stronger long-range interactions the data are consistent with $\lambda=d/2=1$.
  For very long-ranged interactions, strong finite-size effects are observed.
  We discuss whether such finite-size effects could be misinterpreted phenomenologically as sub-aging. 
\end{abstract}

\maketitle
The time evolution of complex systems after a quench from a disordered state at high temperature to a low temperature where the equilibrium state has a non-zero order parameter is characterized by dynamical scaling laws describing coarsening and aging phenomena \cite{Bouchaud_book,bray2002theory,puri2009kinetics,henkel2010non}.
Understanding this nonequilibrium phase-ordering kinetics is key for predicting structure formation processes in many fields. Applications range from statistical and soft-matter physics at mesoscopic scales \cite{Fisher1988,Mathieu2001,Castillo2002,Lunkenheimer2005,rothlein2006symmetry,*rothlein2007erratum,henkel2012phenomenology,majumder2016PRE,kelling2017local,durang2017,majumder2017SM,christiansen2017JCP} to biology \cite{Costa2005,lou2017linking}, from quantum physics at the nanoscale \cite{schutz1999relaxation,hofmann2014coarsening,jeong2014coarsening,maraga2015aging,mitra2018quantum} to astrophysics \cite{kibble1980some,witten1984cosmic,vilenkin1985cosmic,vilenkin1985cosmic,binney2011galactic} at the cosmic scale.
In many of these systems, long-range interactions play an important role \cite{eyink2006onsager,campa2009statistical,french2010long,levin2014nonequilibrium,campa2014physics,douglas2015quantum,neyenhuis2017observation,zhang2018long}, which are hard to deal with both theoretically and computationally. 
For aging of long-range interacting systems comparatively little is known theoretically, with the notable exception of analytical studies of the long-range spherical model \cite{cannas2001dynamics,baumann2007kinetics}.
\par
In this work we therefore strive to uncover the most distinguishing features of aging of long-range interacting systems when compared to the short-range case.
To avoid distractions from system-specific details as much as possible, we consider the paradigmatic two-dimensional (2D) long-range Ising model (LRIM), with Hamiltonian
\begin{equation}\label{LRIM}
\mathcal{H}=-\frac{1}{2} \sum_i\sum_{j \ne i}J(r_{ij}) s_is_j~\textrm{and}~J(r_{ij})=\frac{1}{r_{ij}^{d+\sigma}}.
\end{equation}
The interaction strength $J(r_{ij})$ depends on the distance $r_{ij}$ between the spins at sites $i$ and $j$ that take values $s_i=\pm 1$.
The exponent $\sigma$ governing the power-law decay enables us to interpolate between the short-range nearest-neighbor Ising model (NNIM) over intermediate-range to extremely long-range interactions, encompassing all interaction patterns encountered in nature.

\par
\begin{figure}
 \includegraphics{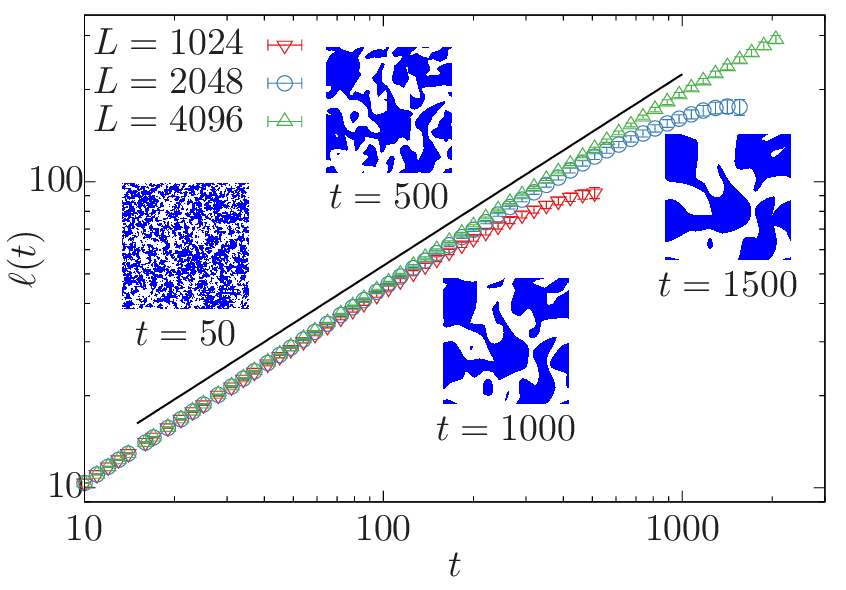}
 \caption{
   Characteristic length scale $\ell(t)$ versus time $t$ for the 2D LRIM with $\sigma=0.6$ on $L \times L$ lattices quenched to $T=0.1T_c$.
   The solid line depicts the theoretical prediction in Eq.~\eqref{Prediction}.
   The snapshots are obtained from a single run for $L=4096$, with spins pointing up marked in blue.}
 \label{length}
\end{figure}

For quenches of the LRIM into the ordered phase with $T<T_c$, the system's long-time behavior is characterized by the existence of a single time-dependent length scale, $\ell(t)$, where for phase-ordering kinetics in \emph{any} dimension it has been predicted that \cite{Bray_only,bray1994growth,Rutenberg1994}
\begin{equation}
 \ell(t)\propto t^{1/z} =
\begin{cases}
 t^{\frac{1}{1+\sigma}} & \sigma < 1 \\
 (t \ln t)^{1/2} & \sigma = 1 \\
 t^{\frac{1}{2}} & \sigma > 1
\end{cases},
\label{Prediction}
\end{equation}
with $z$ denoting the dynamical exponent \footnote{In general, one has $\ell(t) \propto t^{\alpha}$, where in most cases $\alpha=1/z$. However, there exist systems such as fluid mixtures incorporating hydrodynamic effects, where this is not true.}.
In Fig.~\ref{length} we show an illustration with $\sigma=0.6$, where $\ell(t)$ has been extracted as the distance where the equal-time two-point correlation function has decayed to $50\%$ (for details see Supplemental Material).

\par
For a proper understanding of the nonequilibrium process, along with single-time quantities one needs to study multiple-time quantities as well, which provide information about the change in properties of a system with its growing age, i.e., its aging characteristics.
Here, this is probed via the two-time autocorrelation function
\begin{equation}\label{auto_Cr}
  C(t,t_w)=\langle \psi (\vec{r}, t)\psi(\vec{r}, t_w)\rangle
\end{equation}
where $\psi$ is the space- and time-dependent order parameter, and $t_w$ ($\le t$) is the waiting time.
In our case, the order parameter is given as $\psi(\vec{r},t)=s_i(t)$.
Simple aging for quenches to $T<T_c$ is characterized \cite{henkel2010non} by slow dynamics, absence of time-translation invariance and dynamical scaling in the scaling variable $y \equiv t/t_w$.
In general, for large $y$ one expects 
\begin{equation}\label{aging_form_lw}
 C(yt_w,t_w) = f_{C}(y) \xrightarrow{y \rightarrow \infty} f_{C,\infty} y^{-\lambda / z},
\end{equation}
where
$\lambda$ is the autocorrelation exponent.
It was assumed that $t_w \gg t_{\mathrm{micro}}$, and $t-t_w \gg t_{\mathrm{micro}}$, where $t_{\mathrm{micro}}$ is some microscopic reference time scale.

For the NNIM a lower bound $\lambda \geq d/2$ \cite{Fisher1988,yeung1996bounds} exists.
Most simulations \cite{Zannetti_book,henkel2010non,henkel2004,Lorenz2007,Majumder2013PRL,Midya2014} in 2D are compatible with $\lambda \approx 1.25$ and it has been argued that $\lambda \leq 1.25$ \cite{Fisher1988}.
For the LRIM, it is \emph{a priori} unclear if this bound should also apply.
Equation (\ref{Prediction}) suggests that for $\sigma>1$ the non-equilibrium behavior might be in the same universality class as the NNIM.
If that should be the case, the autocorrelation exponent $\lambda \approx 1.25$ is expected.
No prediction for $\lambda$ exists for $\sigma \leq 1$.

\begin{figure}[!]
  \centering
  \includegraphics{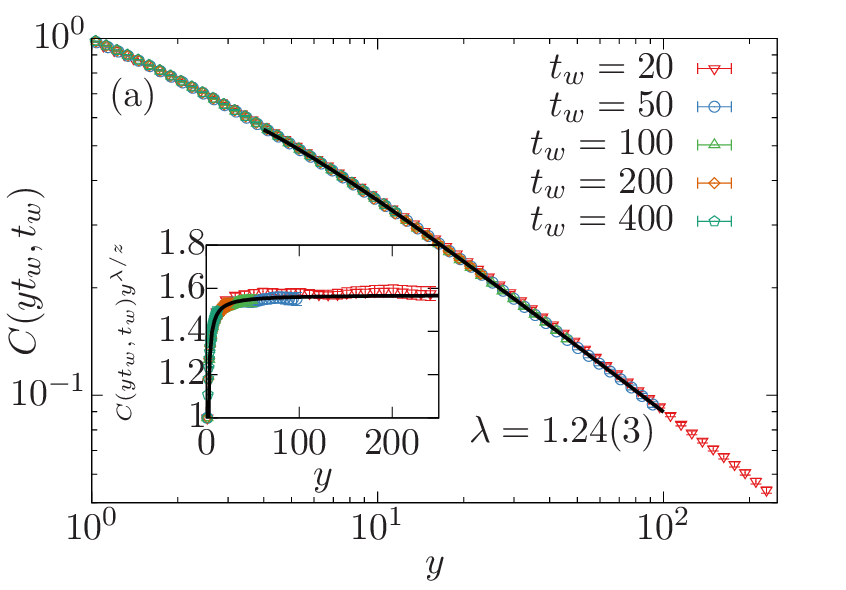}
  \includegraphics{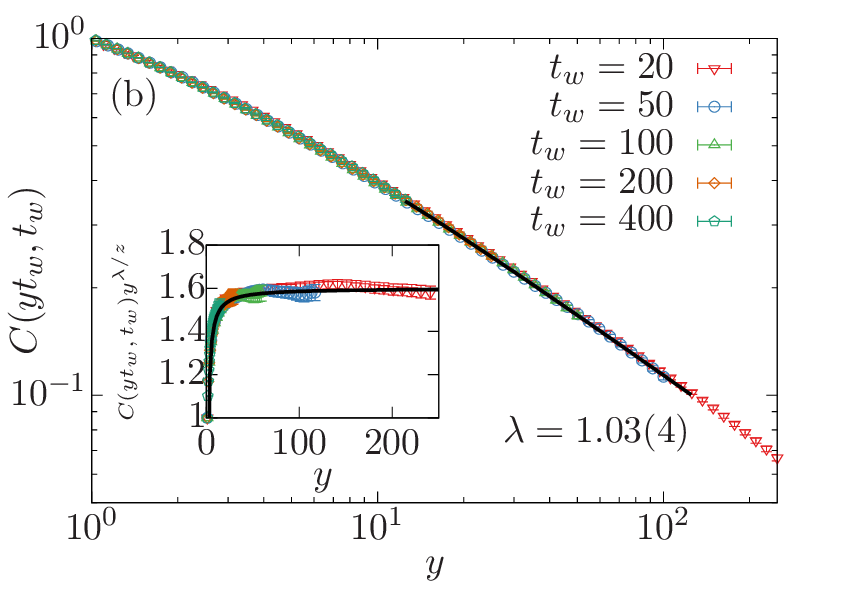}
  \includegraphics{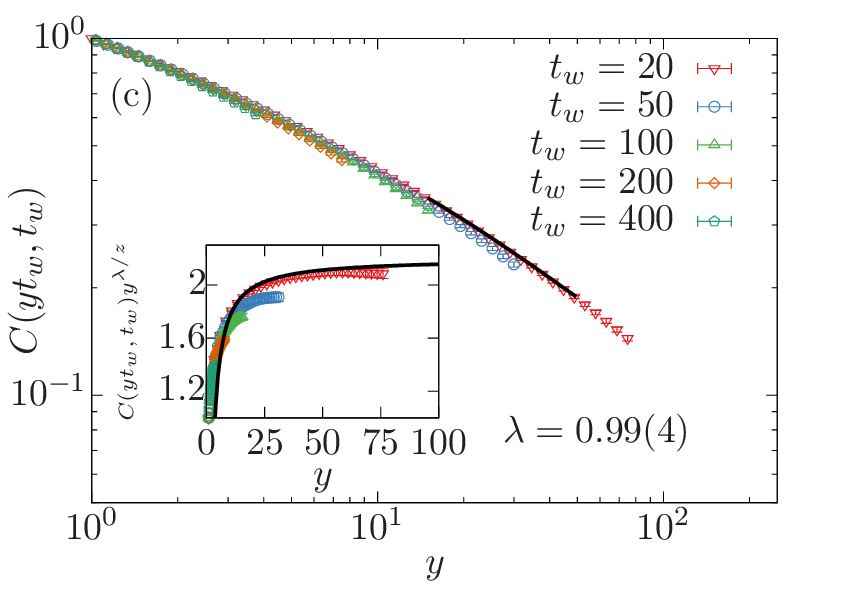}
  \caption{ Double-log plot of the order-parameter autocorrelation function $C(yt_w,t_w$) against the scaling variable $y = t/t_w$ for the 2D LRIM with (a) $\sigma=1.5$ with $L=2048$ and (b) $\sigma=0.8$ and (c) $\sigma=0.6$ with $L=4096$ quenched to $T=0.1T_c$.
    The solid lines are fits using the form (\ref{CorrectionForm}).
    In the insets we plot the same data on a linear scale, but dividing out the asymptotic behavior $y^{-\lambda/z}$.
  }
  \label{Fig_tbytw}
\end{figure}
We study the phase-ordering kinetics of the LRIM on $L \times L$ periodic lattices via Monte Carlo simulations (see Supplemental Material) by quenching to $T=0.1T_c(\sigma)$.
The values of $T_c(\sigma)$ are from recent equilibrium studies of this model focusing on the critical regime \cite{horita2017upper}.
All presented results are averaged over at least $30$ independent realizations.
The error bars are of the order of the size of the data points if not shown.
\par
The long-standing theoretical prediction (\ref{Prediction}) for $\ell(t)$ has only recently been confirmed by us in 2D \cite{christiansen2018} and subsequently in 1D \cite{corberi2019one}.
In Fig.~\ref{length} we plot $\ell(t)$ versus $t$ for $\sigma=0.6$ and $L=1024,2048$ \cite{christiansen2018} and add new data for $L=4096$.
Clear finite-size effects are seen, such that for $t>t_{\times}(L)$ deviations from the infinite system occur.
We estimate the onset of finite-size effects as $t_{\times}(1024)\approx 100-200$ and $t_{\times}(2048) \approx 300-500$.
The exemplary snapshots from a single run illustrate how the emergent structures grow with time.
\par
We now focus on the main part of this work, the two-time correlator $C(yt_w,t_w)$.
When plotted against $t-t_w$ we get curves which relax slower with increasing $t_w$, implying the absence of time-translation invariance (see Supplemental Material).
We start our quantitative analysis for a case for which we expect behavior similar to the NNIM, i.e., the 2D LRIM in the short-range regime with $\sigma=1.5$.
In Fig.~\ref{Fig_tbytw}(a) we test for dynamical scaling by plotting $C(yt_w,t_w)$ for $L=2048$ against $y$ on a log-log scale for different waiting times $t_w$.
The good data collapse onto a master curve $f_C(y)$ clearly validates the simple scaling scenario.
For small $y$, we observe a curvature indicating corrections to the asymptotic power law (\ref{aging_form_lw}) which we assume to be in leading order of the form 
\begin{equation}
  C(yt_w,t_w) =  f_{C,\infty} y^{-\lambda/z}\left(1-\frac{A}{y}\right).
  \label{CorrectionForm}
\end{equation}
This is a generic ansatz, which is known from the exactly solved spherical \cite{henkel2010non,cannas2001dynamics,baumann2007kinetics} and 1D Glauber-Ising \cite{henkel2010non} models.
More generally, this is also known from local scale-invariance (LSI) \cite{picone2004local} for any phase-ordering system with $z=2$, where it can also be shown that $A \geq d-\lambda$ (see Supplemental Material).
When fitting this ansatz to the data points, we first systematically vary the lower $t_{\mathrm{min}}$ and upper $t_{\mathrm{max}}$ boundaries of the fit window.
Out of the resulting $100-200$ fits, we select a particular fit by demanding $\Delta t = t_{\mathrm{max}} - t_{\mathrm{min}}$ to be maximal, under the constraint that the reduced $\chi^2_r$ has no (strong) systematic trend.
Effectively, $t_{\mathrm{min}}$ thus checks down to which $t$ the data is well described by the first-order correction $A/y$ and $t_{\mathrm{max}}$ detects the onset of noticeable finite-size effects.
Since the data are (trivially) correlated in time the value of $\chi^2_r$ has no absolute interpretation, but a comparison of different fitting ranges is still meaningful.
All fits in the region where $\chi^2_r$ has no clear trend only show a systematic variation within $1\%-2\%$ for $\lambda$.
Statistical errors on the fit parameters were estimated from a Jackknife analysis \cite{efron1982jackknife}, i.e., we performed an independent fit for each Jackknife bin (containing all but data from one seed).
For $\sigma=1.5$ we have chosen the data for $t_w=50$ for our analysis, since small deviations from the master curve are visible for $t_w=20$.
This corresponds to a value of $\ell(t_w)=12.51(2)$, which is clearly in the scaling regime.
In this case, the lower bound is $t_{\mathrm{min}}=200$ and the upper bound is $t_{\mathrm{max}}=5000$, where this is the last available data point (i.e., up to this point there are no detectable finite-size effects, see Supplemental Material).
For this fit window, we find $\lambda=1.24(3)$, $A=0.7(1)$, and $f_{C,\infty}=1.57(7)$.
This is perfectly consistent with $\lambda=1.25$ as expected for the NNIM.
The solid line in Fig.~\ref{Fig_tbytw}(a) shows this fit.
In the inset we plot the same data with the asymptotically expected power law $y^{-\lambda/z}$ divided out, which implies a constant behavior in the asymptotic limit.
From the pronounced curvature for small $y$ the $1/y$ correction in (\ref{CorrectionForm}) to the asymptotic power law (\ref{aging_form_lw}) is evident.
The solid line shows again the fit.
\par
Next, we consider the case $\sigma=0.8$.
According to Eq. (\ref{length}), this should be distinct from the short-range universality class.
Our analysis follows the method developed above for $\sigma=1.5$. 
In Fig.~\ref{Fig_tbytw}(b) we show the autocorrelation $C(yt_w,t_w)$ for $L=4096$ as a function of $y$, which collapses onto a master curve for all shown $t_w$.
Here, we use the data with $t_w=20$ for our fits, as this provides the longest possible fitting ranges and $\ell(t_w)=11.90(2)$ is in the scaling regime and compatible with $\ell(t_w)$ used for $\sigma=1.5$.
Fitting the form (\ref{CorrectionForm}) shows no systematic trend in the range from $t_{\mathrm{min}}=250$ to $t_{\mathrm{max}}=2500$, giving estimates of the fit parameters as $\lambda=1.03(4)$, $f_{C,\infty}=1.6(2)$, and $A=0.9(6)$, which is compatible with $\lambda=d/2=1$, the (putative) lower bound on $\lambda$.
In the phase-ordering long-range spherical model, one also finds $\lambda=d/2$ \cite{cannas2001dynamics,baumann2007kinetics}.
The quality of the fit is visually reinforced by the solid lines both in the main plot and the inset.
Here $t_{\mathrm{max}}$ is understood as an estimator for $t_{\times}$, the time where detectable finite-size effects set in.
Data with $t>t_{\times}$ decay stronger than the assumed asymptotic power for all $t_w$ (see Supplemental Material).
In general, these finite-size effects always occur at the same value of $t_{\times}$, thus effectively at different $y$.
From the inset we see a deviation from the trend of the data for $t_w=20$ at $y \approx 150$, which indicates that all data for $t\gtrapprox 3000$ need to be disregarded, compatible with $t_{\times} = 2500$ estimated from the fits.
\par
\begin{figure}
  \includegraphics{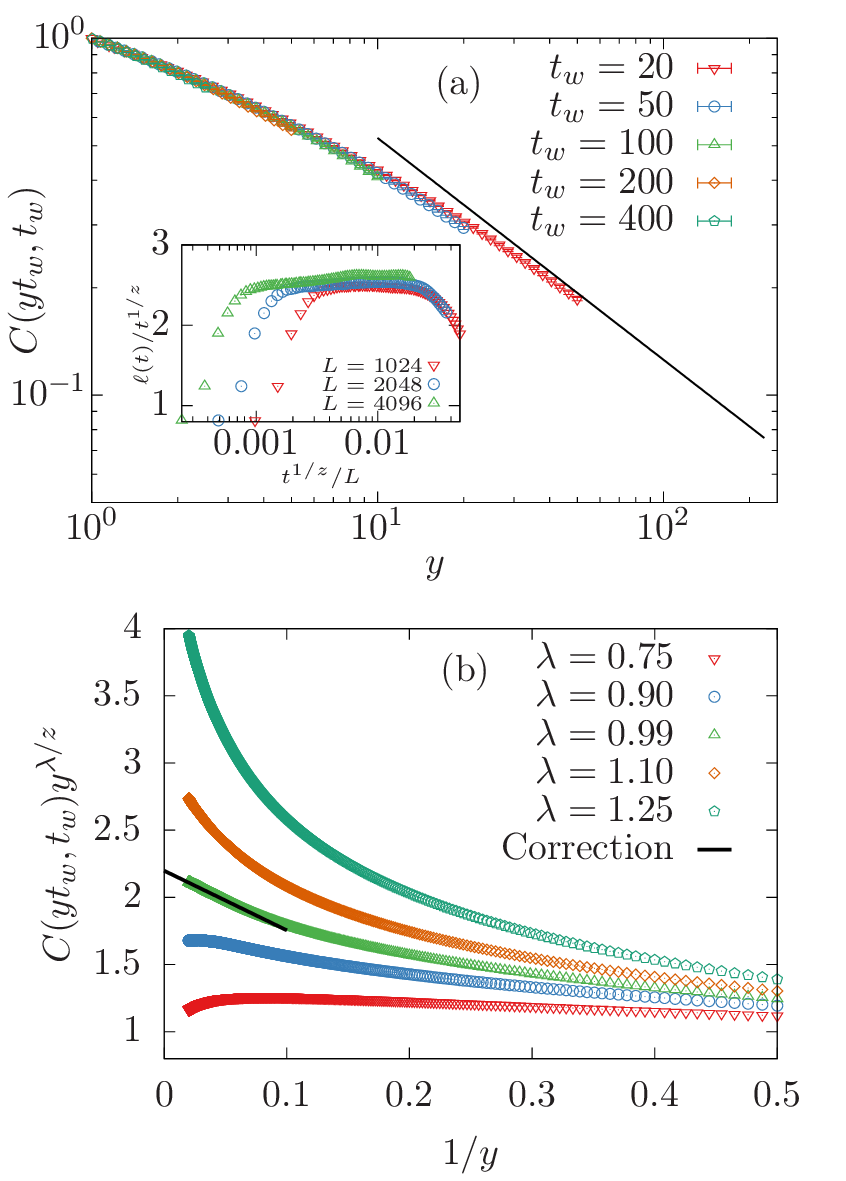}
  \caption{
    (a) $C(yt_w,t_w)$ versus $y$ for $\sigma=0.6$ and $L=4096$, omitting the finite-size affected data for $t>t_{\times}=1000$.
    The straight line shows the asymptotic power law $y^{-\lambda/z}$ with the fitted $\lambda=0.99$.
    The inset shows a finite-size scaling plot of $\ell(t)$ by plotting $\ell(t)/t^{1/z}$ versus $t^{1/z}/L$ for different $L$.
    (b) Plot against $1/y$ of the data for $t_w=20$, dividing out the asymptotically expected behavior of $y^{-\lambda/z}$.
    The assumed value of $\lambda$ is varied and the solid line is the expected behavior assuming correction form (\ref{CorrectionForm}).
  }
  \label{FSSSigma6}
\end{figure}
In contrast, for the data with $\sigma=0.6$ and $L=4096$ shown in Fig.~\ref{Fig_tbytw}(c) one observes two problems: (\emph{i}) there is apparently no completely satisfactory data collapse and (\emph{ii}) there is no pronounced power-law like scaling regime.
Performing fits using (\ref{CorrectionForm}) for $t_w=20$ (with $\ell(t_w)=16.27(3)$) suggests $t_{\mathrm{max}}=1000$, while $t_{\mathrm{min}}=300$ appears suitable as a lower bound.
This gives $\lambda=0.99(4)$, $A=2.0(4)$, and $f_{C,\infty}=2.2(2)$, which is once more consistent with $\lambda=d/2=1$.
However, since $\Delta t$ is rather short and the visual impression of the data collapse is not perfect, we analyze this case in more detail.
\par
Using $t_{\mathrm{max}}=1000$ as the estimate for $t_{\times}$, we replot in Fig.~\ref{FSSSigma6}(a) the data by omitting all points with $t>t_{\times}$, giving a much improved impression of data collapse.
The estimate for $t_{\times}$ is thus crucial for the visual judgment.
To substantiate this, we now estimate the onset of finite-size effects also from the data of $\ell(t)$.
For this, we plot in the inset of Fig.~\ref{FSSSigma6}(a) $\ell(t)/t^{1/z}$ versus $t^{1/z}/L$.
The onset of finite-size effects in this representation is independent of $L$ and happens at the same value of $t^{1/z}/L$.
Here one could read off values of $t_{\times}^{1/z}/L$ between $\approx 0.012$ and $\approx 0.020$, corresponding for $L=4096$ to the relatively wide range $510  \lessapprox t_{\times}  \lessapprox 1150$, compatible with $t_{\times}$ extracted from the trends of $\chi^2_r$.
The straight line in Fig.~\ref{FSSSigma6}(a) shows $y^{-\lambda/z}$ with $\lambda=0.99$ from the fit.
In Fig.~\ref{FSSSigma6}(b) we plot $C(yt_w,t_w)y^{\lambda/z}$ against $1/y$, emphasizing the asymptotic behavior when compared to the insets of Fig.~\ref{Fig_tbytw}:
For the correct $\lambda$ the data should approach $f_{C,\infty}$ linearly as $1/y \rightarrow 0$.
If $\lambda$ is too large, the data diverge for $1/y \rightarrow 0$, whereas for $\lambda$ too small, a downward tendency is expected.
From the plot, we observe for $\lambda<0.99$ this downward tendency, while for $\lambda>0.99$ the curves have increasing slope.
For $\lambda=0.99$ the approach to $1/y \rightarrow 0$ is indeed linear with constant slope over a significant range, which is verified by the solid line as obtained from the fit.
\par
\begin{figure}
  \includegraphics{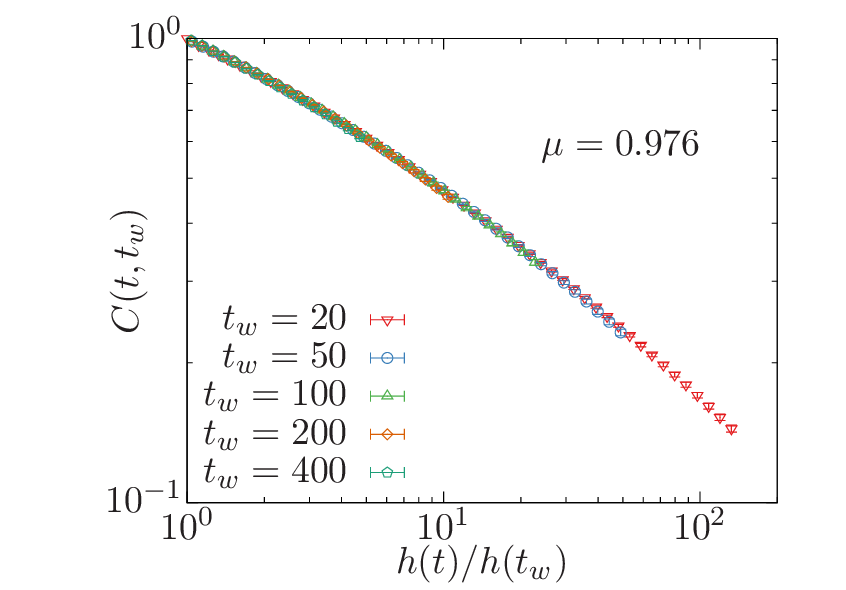}
  \caption{Untruncated $C(yt_w,t_w)$ for $\sigma=0.6$ and $L=4096$ plotted against scaling variable $h(t)/h(t_w)$ with $\mu=0.976$, indicating sub-aging.
  }
  \label{SubagingPlot}
\end{figure}
We should point out, however, that our current data for $\sigma=0.6$ would also be compatible with the alternative interpretation of exhibiting sub-aging behavior.
In this scenario one considers the scaling ansatz 
\begin{equation}
  C(t,t_w) = \tilde{f}_C\left(\frac{h(t)}{h(t_w)}\right),
  \label{FitFormCorr}
\end{equation}
with $h(t) \equiv  \exp\left((t^{1-\mu} -1)/(1-\mu) \right)$, where the parameter $\mu$ characterizes the deviation from simple scaling which is recovered in the limit $\mu \rightarrow 1$.
Sub-aging with $\mu<1$ has been encountered many times in analytical \cite{berthier2000sub,durang2017}, numerical \cite{kim2003nonequilibrium,grempel2004off,pluchino2004glassy,tamarit2005relaxation}, and experimental investigations, see Ref.~\cite{henkel2010non} for a list of examples.
In Fig.~\ref{SubagingPlot} we show the scaling with respect to $h(t)/h(t_w)$ for the untruncated data, where $\mu \approx 0.976$ provides the best data collapse.
Compared to Fig.~\ref{Fig_tbytw}(c) showing the same data, the data collapse is greatly improved.
Of course, sub-aging introduces one additional tunable parameter and as such one would always expect better data collapse.
On the other hand, we have provided rather strong evidence that the (slight) downward bending of the curves in Fig.~\ref{Fig_tbytw}(c) for large $y$ is caused by finite-size effects [cf., Fig.~S2(c) of the Supplemental Material].
Assuming asymptotically $C(t,t_w) \rightarrow [h(t)/h(t_w)]^{-\tilde{\lambda}/z}$, where $\tilde{\lambda}$ is a modified autocorrelation exponent, one has that for large $y$ (or $t$ for fixed $t_w$) the sub-aging ansatz (\ref{FitFormCorr}) decays faster than any power law, i.e., proportional to $\exp \left(-\frac{\tilde{\lambda}/z}{1-\mu} t^{1-\mu} \right)$.
When plotted as a function of $y$ as in Fig.~\ref{Fig_tbytw}(c), this thus models a downward bending suggesting that the sub-aging scaling collapse just looks so good because this ansatz effectively ``compensates'' the finite-size effects.
Only on the basis of the present data for $\sigma=0.6$ on lattices up to $4096^2$ it is, however, not possible to clearly favor one of the two alternative scaling scenarios. 
Based on our results for the other values of $\sigma$ where we have clear evidence for simple aging, we side with the interpretation of simple aging also for $\sigma=0.6$.
The Supplemental Material presents alternatively the scaling behavior with respect to $t/t_w^{\mu}$, a simpler phenomenological form often used to probe for sub-aging.

To conclude, we have performed the first numerical investigation of aging in long-range systems by systematically tuning the interaction using the paradigmatic two-dimensional long-range Ising model.
We find for all $\sigma$ simple aging, where for $\sigma=0.6$ it is shown that strong finite-size effects may be misinterpreted as sub-aging.
The autocorrelation exponent is consistent with $\lambda=d/2=1$ for $\sigma<1$ and with $\lambda=1.25$ for $\sigma>1$.
This implies that the transition between the short-range and long-range 2D Ising universality classes occurs at a different value of $\sigma$ than it does either {\em at} the critical point or else in equilibrium.
The conjecture $\lambda=d/2$ is consistent with known results:
For the 1D LRIM at $T=0$ one finds $\lambda=0.5$ for $\sigma<1$ \cite{corberi2019universality} and the phase-ordering long-range spherical model has $\lambda=d/2$ independently of $\sigma$ \cite{cannas2001dynamics,baumann2007kinetics}.
\par
An open and interesting question is, how this transition in $\lambda$ happens, i.e., whether it is smooth or is characterized by a jump.
To answer this, even larger systems would have to be simulated, which is out of scope for the time being.
The more involved case of binary mixtures, i.e., a conserved order parameter setting, is enticing as a next step \footnote{Work in progress}.
Crucial is also the investigation of aging in other models with long-range interactions, as this could shed new light on our understanding of aging in liquid crystals \cite{singh2014ordering}, active systems \cite{steimel2016emergent}, or strongly interacting quantum many-body systems \cite{blass2018quantum}.
\par
\begin{acknowledgments}
This project was funded by the Deutsche Forschungsgemeinschaft (DFG, German Research Foundation) under Grant Nos.~JA 483/33-1 and 189\,853\,844 -- SFB/TRR 102 (Project B04), and the Deutsch-Franz\"osische Hochschule (DFH-UFA) through the Doctoral College ``$\mathbb{L}^4$'' under Grant No.~CDFA-02-07. We further acknowledge support by the Leipzig Graduate School of Natural Sciences ``BuildMoNa''. MH thanks the MPI-PKS Dresden for warm hospitality. 
\end{acknowledgments}

\end{document}


\title{Supplemental Material: Aging in the long-range Ising model}
\author{Henrik Christiansen}
\email{henrik.christiansen@itp.uni-leipzig.de}
\affiliation{Institut f\"ur Theoretische Physik, Universit\"at Leipzig, IPF 231101, 04081 Leipzig, Germany}
\author{Suman Majumder}
\email{suman.majumder@itp.uni-leipzig.de}
\affiliation{Institut f\"ur Theoretische Physik, Universit\"at Leipzig, IPF 231101, 04081 Leipzig, Germany}
\author{Malte Henkel}
\email{malte.henkel@univ-lorraine.fr}
\affiliation{Laboratoire de Physique et Chimie Th\'eoriques (CNRS UMR 7019), Universit\'e de Lorraine Nancy, 54506 Vand{\oe}uvre-l\`es-Nancy Cedex, France}
\affiliation{Centro de F\'isica Te\'orica e Computacional, Universidade de Lisboa, 1749-016 Lisboa, Portugal}
\author{Wolfhard Janke}
\email{wolfhard.janke@itp.uni-leipzig.de}
\affiliation{Institut f\"ur Theoretische Physik, Universit\"at Leipzig, IPF 231101, 04081 Leipzig, Germany}
\date{\today}

\maketitle
  
\section{Methods}
For the Monte Carlo (MC) simulation of the LRIM given by the Hamiltonian
\begin{equation}\label{LRIM}
\mathcal{H}=-\frac{1}{2}\sum_i\sum_{j\neq i}J(r_{ij}) s_is_j,~\textrm{with}~J(r_{ij})=\frac{1}{r_{ij}^{d+\sigma}},
\end{equation}
we introduce the kinetics via single-spin flips.  A randomly chosen spin is flipped according to the standard Metropolis update with probability $\min \left[1,\exp(-\Delta E/T)\right]$, with the Boltzmann constant $k_B$ set to unity. Here, $T$ is the temperature and $\Delta E$ is the change in energy before and after the flip. $N=L^d$ (where $L$ is the linear size of a hyper-cubic lattice) such attempts constitute one MC sweep, setting the time scale. Obviously, for the LRIM the calculation of the energy change is the rate limiting step, as it involves all the spins in the considered lattice. However, following our recent approach of storing the effective field for each spin and updating it only when a spin flip is accepted makes such simulation significantly faster \cite{christiansen2018}. Furthermore, to allow for simulations of system size up to $L=4096$ in $d=2$ dimensions, this update was parallelized using the shared-memory API OpenMP framework. Since systems with long-range interaction suffer severely from finite-size effects we additionally use Ewald summation \cite{ewald1921berechnung,horita2017upper,flores2017cluster,ConferenceChristiansen} to implement periodic boundary conditions and thereby to increase the effective system size. An effective $J_{ij} \equiv J(r_{ij})$ is calculated once at the beginning of the simulation. 
\par
As an initial configuration at high temperature, we chose a square lattice with randomly distributed equal proportion of up and down spins. We chose $T=0.1T_c$ as the quench temperature, where we extract $T_c$ from the data presented in Ref.\ \cite{horita2017upper}. Using the scaling relation $C(r,t)\equiv \tilde{C}\left[r/\ell(t)\right]$ for the equal-time two-point correlation function $C(r,t)=\langle s_i s_j \rangle - \langle s_i \rangle \langle s_j \rangle$ one can estimate the characteristic length scale $\ell(t)$ from the decay of $C(r,t)$ as intersection with a constant value where here we choose $C \left[r=\ell(t),t\right]=0.5$. All considered quantities such as $\ell(t)$ and $C(t,t_w)$ are averages over independent time evolutions, indicated, e.g., in Eq.\ (3) of the main article by $\langle \ldots \rangle$. The presented results are averaged over $50$ independent runs for $L \leq 2048$ and $30$ for $L=4096$ (using different random number seeds).




\section{Illustration of the Loss of Time-Translational Invariance}
\begin{figure}
  \centering
  \includegraphics{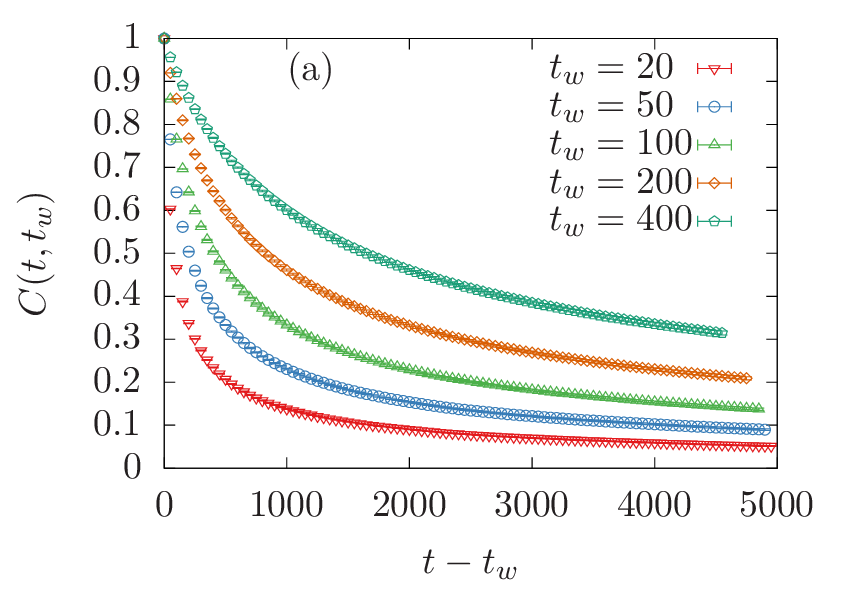}
  \includegraphics{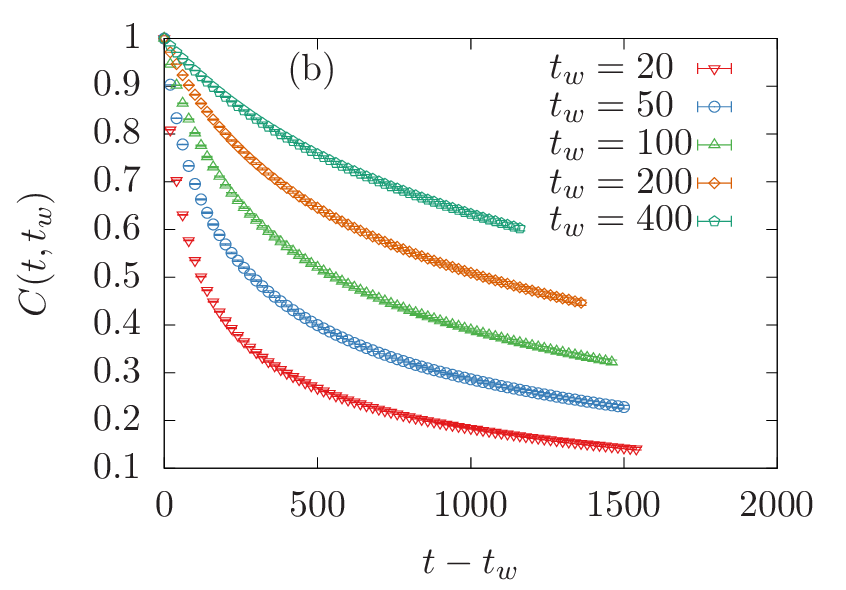}
  \caption{Two-time correlator $C(t,t_w)$ plotted against $t-t_w$, illustrating the loss of time-translational invariance for (a) $\sigma=1.5$ and $L=2048$ and (b) $\sigma=0.6$ and $L=4096$.}
  \label{fig:time_trans_loss}
\end{figure}

Figure~\ref{fig:time_trans_loss} shows the two-time correlator $C(t,t_w)$ versus $t-t_w$, explicitly demonstrating the loss of time-translational invariance during coarsening.
The data for larger $t_w$ decay slower, i.e., the older the system is at the waiting time $t_w$, the longer in terms of $t$ it needs to decorrelate.

\section{Finite-size effects of the Autocorrelation Function}
\begin{figure}
  \centering
  \includegraphics{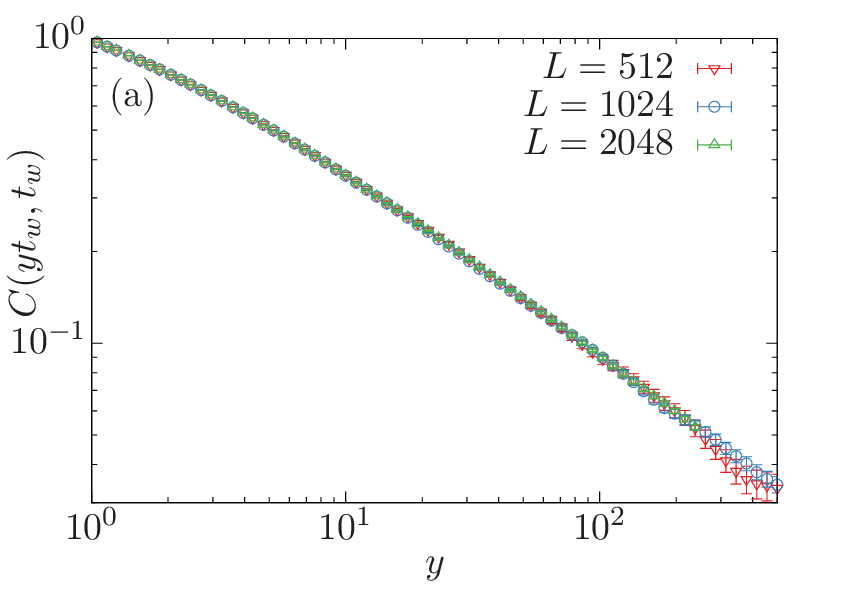}
  \includegraphics{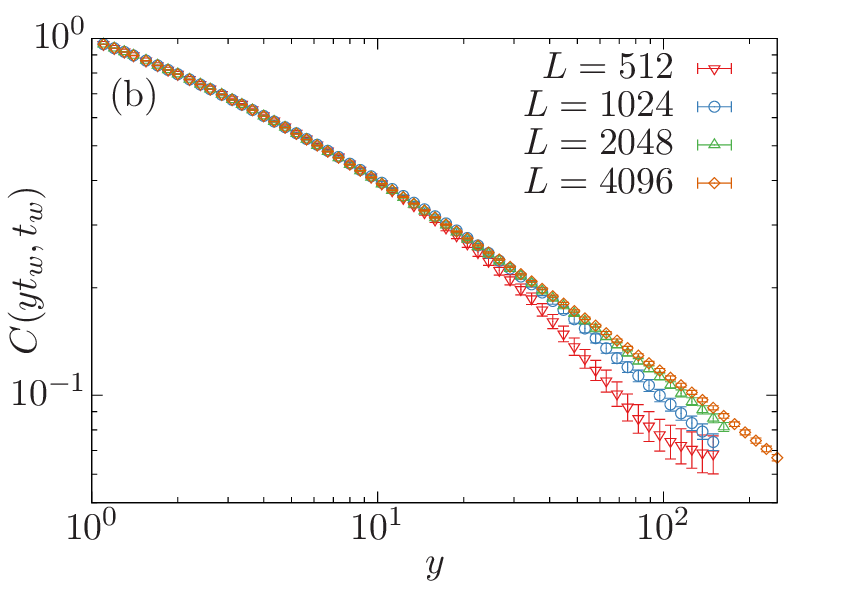}
  \includegraphics{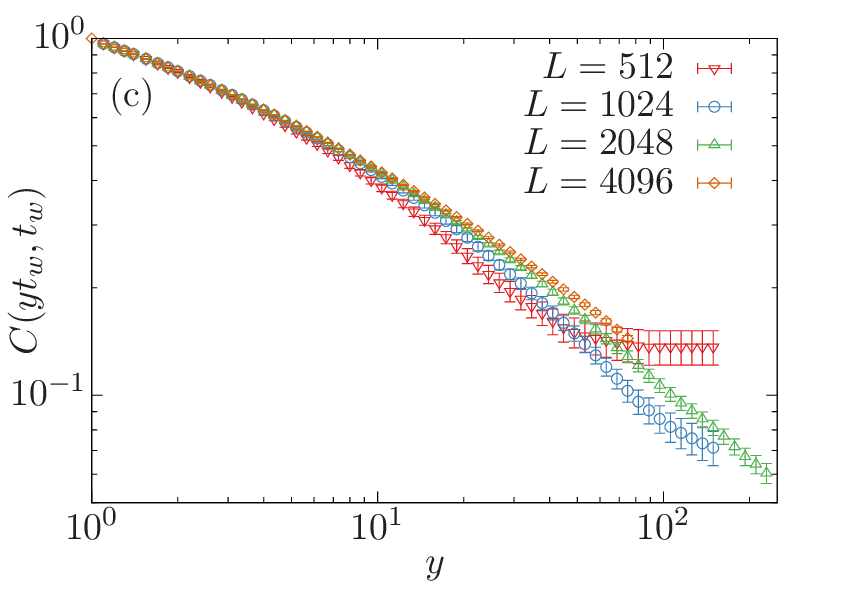}
  \caption{As an illustration of finite-size effects, we show $C(yt_w,t_w)$ for fixed $t_w=20$ by varying the system size $L$ for (a) $\sigma=1.5$, (b) $\sigma=0.8$, and (c) $\sigma=0.6$.}
  \label{fig:FS}
\end{figure}
In Fig.~\ref{fig:FS} we show $C(yt_w,t_w)$ versus $y$ for (a) $\sigma=1.5$, (b) $\sigma=0.8$, and (c) $\sigma=0.6$ with fixed $t_w=20$ and varying $L$.
For $\sigma=1.5$ the data show the bulk behavior over a large $y$ range, and only for $L=512$ the data deviate by bending down at $y\approx 200$.
The available data for $L=1024$ and $L=2048$ do not deviate, i.e., there are no detectable finite-size effects.
For $\sigma=0.8$ and $\sigma=0.6$, the data for both $L=512$ and $L=1024$ undershoot from the bulk curve.
This happens at larger $y$, the larger $L$.
For $L=2048$ this effect is hence less pronounced and for $L=4096$ it can only be anticipated from these plots.
Finally, because eventually the system reaches a configuration with spontaneous magnetization $m_{\mathrm{eq}}(T)$, the overlap and thereby autocorrelation function approaches a constant.
Note that the data for smaller systems even cross the data of the bigger systems.
This effectively limits the extent to which the data can undershoot from the bulk behavior for smaller $y$.

\begin{figure*}
  \includegraphics{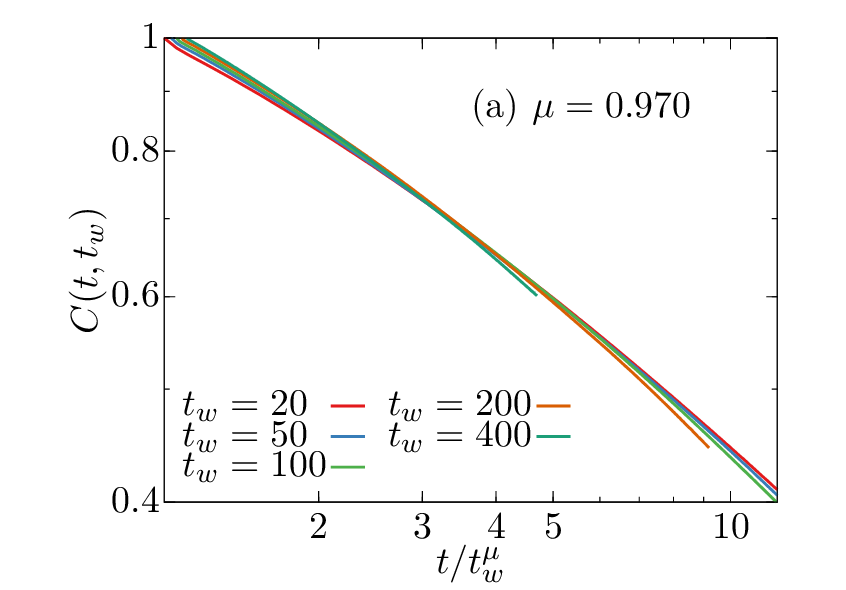}  \includegraphics{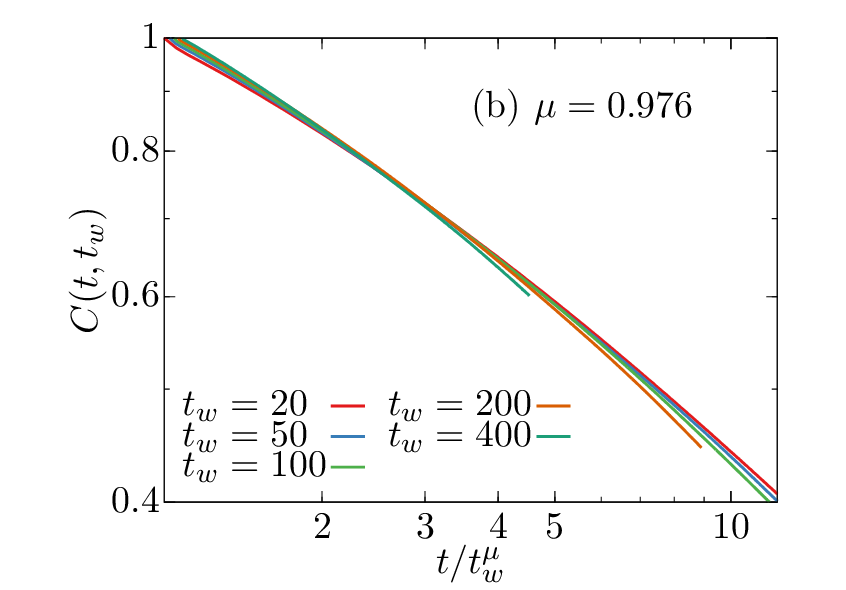}
  \includegraphics{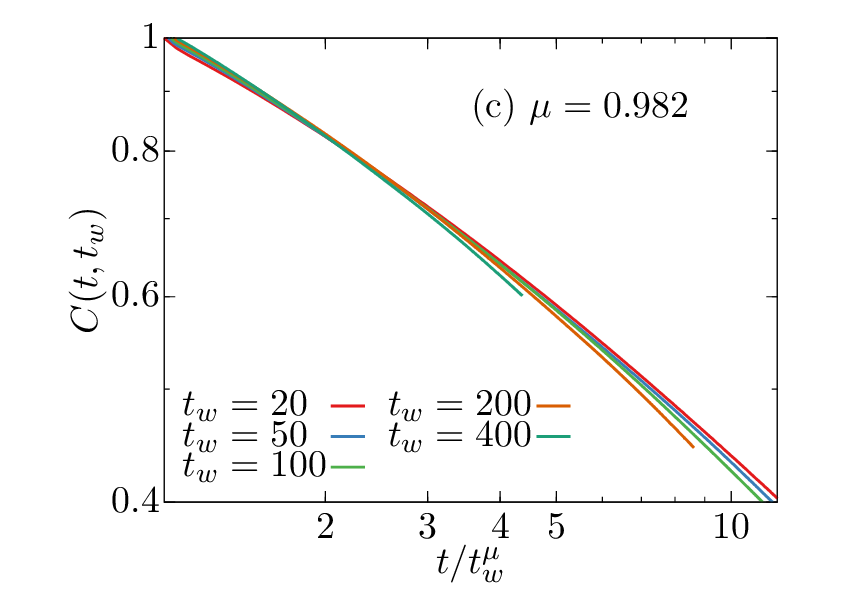}  \includegraphics{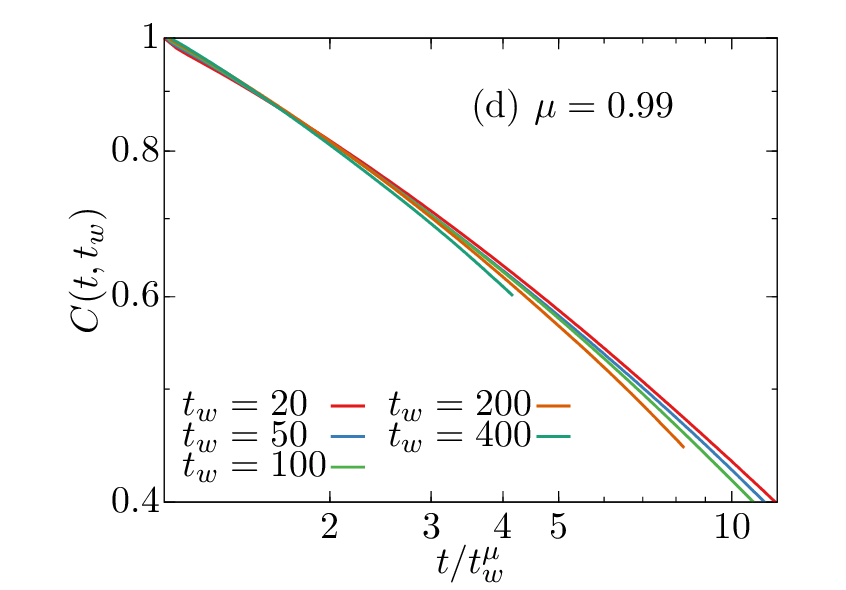}
  \caption{Plots of $C(t,t_w)$ against $t/t_w^{\mu}$ for different values of $\mu$ mentioned in the figure. The data presented is for $\sigma=0.6$ and $L=4096$.}
  \label{OtherSubaging}
\end{figure*}

\section{Alternative Form of Sub-aging}
Instead of using the analytically derived form of sub-aging with $h(t)$ as defined in the main article, one may use the more phenomenological form of $t/t_w^{\mu}$ (or $\ell(t)/\ell(t_w)^{\mu}$) to modify the scaling variable. In Fig.~\ref{OtherSubaging} we present  $C(t,t_w)$ vs. $t/t_w^{\mu}$ for $\sigma=0.6$ and $L=4096$ with $\mu=0.970,0.976,0.982,$ and $0.99$. Compared to using $h(t)$ the data collapse is worse and one effectively only shifts the crossing point of data for different $t_w$. This approach does thus not lead to better collapse.

\section{Two-time autocorrelators from local scale-invariance with $z=2$} 

\newcommand{\D}{{\rm d}}                

According to local scale-invariance \cite{henkel1994lsi,henkel2010non,henkel2017tuteur} the generic dynamical scaling which arises especially in aging systems far from equilibrium can be extended to a larger group of dynamical symmetries.
For the phase-ordering kinetics of systems with short-ranged interactions, it is known that the dynamical exponent $z=2$ \cite{bray1994growth,bray2002theory}.
Then the Schr\"odinger group, which arises as dynamical symmetry of the free diffusion equation, is an example of an extended dynamical symmetry \cite{henkel1994lsi}. 
Numerous systems which physically realize Schr\"odinger invariance have been found, most notably phase-ordering kinetics in short-ranged Ising models in $d=1,2,3$ dimensions, see \cite{henkel2010non,henkel2017tuteur} and references therein.
Here we discuss how the requirement of Schr\"odinger invariance restricts the two-time autocorrelator in phase-ordering kinetics. 

Physically, it is the two-time or multi-time response functions which transform co-variantly under local scale-transformations. 
Turning to the two-time autocorrelator $C(t,t_w)$, after a quench to $T<T_c$ from a fully disordered initial state, it can be expressed as \cite{picone2004local}
\begin{equation}
C(t,t_w) = \frac{a_0}{2} \int_{\mathbb{R}^d} \!\D \vec{R}\: {\cal R}^{(3)}(t,t_w,t_{\rm micro};\vec{R})
\end{equation}
where ${\cal R}^{(3)}$ is a three-point response function which in the context of Janssen-de Dominicis theory could be expressed as an average $\left\langle\psi(t,\vec{0})\psi(t_w,\vec{0})\widetilde{\psi}(t_{\rm micro};\vec{R})^2\right\rangle$ involving the order parameter $\psi$ and the conjugate response operator $\widetilde{\psi}$. 
The form of that three-point response in turn is fixed up to a scaling function \cite{henkel1994lsi}.
Furthermore, $t_{\rm micro}$ is a microscopic time scale and the amplitude $a_0$ measures the width of the initial correlator.
Since for phase-ordering kinetics the temperature $T$ is an irrelevant variable \cite{bray2002theory}, the thermal heat bath merely furnishes corrections to scaling.
In the dynamical scaling regime, the autocorrelator of phase-ordering kinetics can then be written as follows \cite{picone2004local},
\begin{eqnarray}
C(y t_w,t_w) &=& f_C(y) \:=\: y^{\lambda/2} (y-1)^{-\lambda} \mathbf{\Psi}\left(\frac{y+1}{y-1}\right),   \label{lsi-supp-3} \\   
\mathbf{\Psi}(w) &=& \int_{\mathbb{R}^d} \!\D\vec{R}\: \exp\left(-\frac{{\cal M}w}{2} \vec{R}^2 \right) \Psi\left(\frac{\cal M}{2}\vec{R}^2\right) ~~~~~~ \label{lsi-supp-4}
\end{eqnarray}
where the undetermined scaling function $\Psi(\varrho)$ comes from the three-point response function mentioned above.
Because of the known asymptotics $f_C(y)\sim y^{-\lambda/2}$ for $y\to\infty$, it follows that $\mathbf{\Psi}(1)$ exists and is finite. Denoting by $S_d$ the surface of the hypersphere in $d$ dimensions, 
Eq.~(\ref{lsi-supp-4}) is re-written in spherical coordinates as
\begin{equation} \label{lsi-supp-5} 
\mathbf{\Psi}(w) = \frac{S_d}{2}\left(\frac{2}{\cal M}\right)^{d/2}  \int_0^{\infty} \!\D\varrho\:~ e^{-w\varrho}\: \varrho^{(d-2)/2} \Psi\left( \varrho\right)
\end{equation}
and we also made explicit the non-universal metric factor $\cal M$. We recognize from this that $\mathbf{\Psi}$ is the Laplace transform of the function $\varrho^{(d-2)/2} \Psi(\varrho)$. 
Since it is well-known that a Laplace transform is infinitely often differentiable wherever it is defined, we can asymptotically expand in $y$ [or equivalently around $w=1$, see (\ref{lsi-supp-3})] and find (the prime denotes the derivative)


\begin{eqnarray}
f_C(y) &=& \mathbf{\Psi}(1)\, y^{-\lambda/2} \left[ 1 + \left( \lambda + 2 \frac{\mathbf{\Psi}'(1)}{\mathbf{\Psi}(1)} \right) \frac{1}{y} + \ldots \right] \nonumber \\
&=& f_{C,\infty}\, y^{-\lambda/2} \left( 1 - \frac{A}{y} + \ldots \right) \label{lsi-supp-6}
\end{eqnarray}
where we identified the constant $A$. This is the form (5) used in the text.

In order to estimate the amplitude $A$, we require some more input on the scaling function $\Psi(\varrho)$ in (\ref{lsi-supp-4}). 
First, we assume that for $\varrho\to\infty$, $\Psi(\varrho)$ grows more slowly than exponentially which is consistent with $\mathbf{\Psi}(1)$ being finite. 
Second, we recall that for $\varrho\to 0$ consistency with the asymptotic scaling of $f_C(y)$ requires that $\Psi(\varrho)\sim \varrho^{\lambda-d/2}$ \cite{picone2004local}. Because of the known bound
$\lambda\geq d/2$ \cite{Fisher1988,yeung1996bounds}, $\Psi(\varrho)$ increases when $\varrho\ll 1$. We strengthen this to the requirement $\Psi'(\varrho)\geq 0$ also when $\varrho$ is finite.  
Next, the integral representation (\ref{lsi-supp-5}) will become useful, via the following estimate 
\begin{eqnarray}
\lefteqn{ \int_0^{\infty} \!\D\varrho\:~ e^{-w\varrho}\: \varrho^{d/2} \Psi(\varrho) 
=  -\frac{1}{w}\int_0^{\infty} \!\D\varrho\:  \frac{\D}{\D \varrho} \left( e^{-w\varrho}\right)  \varrho^{d/2} \Psi(\varrho) }
\nonumber \\
&=& - \underbrace{\left[ \left. \frac{e^{-w\varrho}}{w}\: \varrho^{d/2} \Psi(\varrho) \right|_0^{\infty} \right]}_{=0} 
+  \int_0^{\infty} \!\D\varrho\: \frac{e^{-w\varrho}}{w} \frac{\D}{\D\varrho} \left[ \varrho^{d/2} \Psi(\varrho) \right] 
\nonumber \\
&=&  \frac{1}{w} \int_0^{\infty} \!\D\varrho\: e^{-w\varrho} \left[ \frac{d}{2} \varrho^{(d-2)/2} \Psi(\varrho) + \varrho^{d/2} \underbrace{~\Psi'(\varrho)~}_{\geq 0} \right]
\nonumber \\
&\geq & \frac{1}{w} \frac{d}{2}  \int_0^{\infty} \!\D\varrho\: e^{-w\varrho}  \varrho^{(d-2)/2} \Psi(\varrho)
\end{eqnarray}
where the two assumptions made on $\Psi(\varrho)$ were used explicitly and also for the estimation of the boundary terms after partial integration.  With (\ref{lsi-supp-5}) we have
\begin{eqnarray}
\frac{\mathbf{\Psi}'(w)}{\mathbf{\Psi}(w)} 
= - \frac{\int_0^{\infty} \!\D\varrho\: e^{-w\varrho} \varrho^{d/2} \Psi(\varrho)}{\int_0^{\infty} \!\D\varrho\: e^{-w\varrho} \varrho^{d/2-1} \Psi(\varrho)} 
\leq -\frac{1}{w} \frac{d}{2}. ~~~
\end{eqnarray}
Setting $w=1$, we then have the bound $\mathbf{\Psi}'(1)/\mathbf{\Psi}(1)\leq -d/2$. For the scaling function $f_C(y)$ of (\ref{lsi-supp-6}), this gives
\begin{equation}
f_C(y) \leq \mathbf{\Psi}(1) y^{-\lambda_C/2} \left[ 1 - \left( 2 \frac{d}{2} -\lambda \right)\frac{1}{y} + \ldots\right].
\end{equation}
This upper bound on $f_C(y)$ gives a lower bound on the amplitude in (\ref{lsi-supp-6}) 
\begin{equation} \label{lsi-supp-10}
A\geq d-\lambda. 
\end{equation} 

Indeed, it was argued long ago by Fisher and Huse \cite{Fisher1988} that $\lambda\leq d$. In models which respect this bound, (\ref{lsi-supp-10}) implies that $A\geq 0$. 
The validity of this Fisher-Huse bound was discussed in detail for phase-ordering systems \cite{Majumdar95a}. 
However, for phase-separating model-B dynamics, this Fisher-Huse bound does not hold \cite{yeung1996bounds,Brown99a}.

Equation (\ref{lsi-supp-6}), along with (\ref{lsi-supp-10}), is reproduced in several exactly solvable models of phase-ordering with nearest-neighbour interactions and $z=2$, see \cite{henkel2010non} for details.

For the 1D Glauber-Ising model at $T=0$, we have  
\begin{eqnarray}
f_C(y) &=&\frac{2}{\pi}\arctan \sqrt{\frac{2}{y-1}\,} \nonumber \\
         &\simeq& \frac{\sqrt{8}}{\pi}y^{-1/2}\left(1-\frac{1}{6}\frac{1}{y}+{\rm O}(y^{-2})\right).
\end{eqnarray}
Since $\lambda=1$,  the bound (\ref{lsi-supp-10}) $A\geq 0$ is consistent with the exact result $A=1/6$.

For the spherical model in $d>2$ dimensions and quenched to $T<T_c$, we have
\begin{eqnarray} 
f_C(y) &=& m_{\rm eq}^2 \left[ 2 y^{1/2}/(y+1)\right]^{d/2} \nonumber \\
          &\simeq&  m_{\rm eq}^2 2^{d/2}y^{-d/4}\left(1-\frac{d}{2}\frac{1}{y}+{\rm O}(y^{-2})\right) ~~~
\end{eqnarray}
with the equilibrium magnetization $m_{\rm eq}^2 = 1 - T/T_c$. 
Since $\lambda=d/2$, the bound (\ref{lsi-supp-10}) $A\geq d/2$ coincides with the exact result $A=d/2$.

Equation (\ref{lsi-supp-6}) can also be used as an ansatz for the spherical model with long-ranged interactions.
There is a phase transition in the long-range universality class at a non-vanishing $T_c$ provided $0<\sigma<\min(d,2)$ and $z=\sigma$.
The scaling function of the two-time autocorrelator is
\begin{eqnarray} 
f_C(y) &=& m_{\rm eq}^2 \left[ 2 y^{1/2}/(y+1)\right]^{d/\sigma} \nonumber \\
          &\simeq&  m_{\rm eq}^2 2^{d/\sigma}y^{-d/(2\sigma)}\left(1-\frac{d}{\sigma}\frac{1}{y}+{\rm O}(y^{-2})\right). ~~~~~~
\end{eqnarray}
Hence, $\lambda=d/2$ is $\sigma$-independent and we note that once more $A>0$. 

\bibliography{bibtex.bib}